\documentstyle{article}
\def\theequation{\arabic{section}.\arabic{equation}}

\begin{document}

\vspace*{1cm}

\begin{center}

{\Large  GENERAL SOLUTION OF QUANTUM MASTER EQUATION

\vspace{1cm}

              IN FINITE--DIMENSIONAL CASE.}

\end{center}

\vspace{1.5cm}

\begin{center}
{\large I.A. Batalin and I.V. Tyutin}
\end{center}

\begin{center}
{\it I.E. Tamm Theory Division \\
P.N.Lebedev Physical Institute  \\
Russian Academy of Sciences\\
53 Leninsky Prospect \\
Moscow 117924, Russia}
\end{center}

\vspace{4cm}

\centerline{\bf Abstract}

\noindent
The general solution to the quantum master equation (and its $Sp(2)$ symmetric
counterpart) is constructed explicitly in case of finite number of variables.
It is shown that the finite--dimensional solution is physically trivial and
thus can not be extended directly to cover the case of a local field theory.
In this way we conclude that the locality condition plays an important role
by making it possible to obtain nontrivial physical results when quantizing
gauge field theories on the basis of field--antifield formalism.

\newpage

\section{Introduction}

When quantizing gauge theories directly within Lagrangian formalism
\cite{1} -- \cite{5} one usually proceeds by solving the quantum master
equation

\begin{equation}\label{1.1}
\Delta e^{{i\over\hbar}S}=0
\end{equation}
or its $Sp(2)$ counterparts \cite{6} -- \cite{9}
\begin{equation}\label{1.2}
\bar{\Delta}^a e^{{i\over\hbar}S}=0, \quad a=1,2,
\end{equation}
where $\Delta$ and $\bar{\Delta}^a$ are Fermionic nilpotent operators:

\begin{equation}\label{1.3}
\Delta^2=0,\quad\bar{\Delta}^a\bar{\Delta}^b+\bar{\Delta}^b\bar{\Delta}^a=0.
\end{equation}
The operators $\Delta$ and $\bar{\Delta}^a$ are local second-order differential
operators of the following  characteristic structure:

\begin{equation}\label{1.4}
\Delta=\int dx(-1)^{\varepsilon_A}{\delta\over\delta\Phi^A(x)}
{\delta\over\delta\Phi^*_A(x)}.
\end{equation}

When studying various properties of solutions to the quantum master equation
(arbitrariness of a solution, gauge independence and reparametrization
invariance of physical quantities, structure of renormalization) one considers
the matter in the same manner as if one dealt with the theory of finite
number of variables, in which the operator $\Delta$ were of the form

\begin{equation}\label{1.5}
\Delta=\sum_{A=1}^N(-1)^{\varepsilon_A}{\partial\over\partial\Phi^A}
{\partial\over\partial\Phi^*_A}.
\end{equation}
However, when extending the results, obtained in a finite--dimensional case,
to a local field theory, one should be very careful (for a review of some
results on local gauge field theory see refs. \cite{10}, \cite{11} and
references therein).  This is because of the fact that various
transformations like a change of integration variables or an inversion of
matrices (operators), which are sensible in a finite-dimensional case, may
appear to be inadmissible when extending them to the case of a local field
theory.

For example, in the case of a system with finite number of Bosonic variables,
the actions

\begin{equation}\label{1.6}
S={1\over2}\Phi^A\Lambda_{AB}\Phi^B
\end{equation}
with a positive-defined matrix $\Lambda_{AB}$  are equivalent (up to an improper change of
variables) to the following canonical action

\begin{equation}\label{1.7}
S={1\over2}\sum_A\Phi^A\Phi^A.
\end{equation}
At the same time, within a local field theory, the actions

\begin{equation}\label{1.8}
S_i={1\over2}\int dx(\partial_\mu\Phi^A\partial_\mu\Phi^A-m^2_i\Phi^A\Phi^A),
\quad i=1,2,
\end{equation}
are nonequivalent for $m_1\neq m_2$.

Another example is as follows. One should frequently deal with classes of
functions determined modulo contributions vanishing at
the extremals.  In a theory with finite number of variables, governed by the
action (\ref{1.6}), the only representatives of the mentioned classes are
constants.  On the other hand, in a local field theory with the action
(\ref{1.8}), nontrivial classes certainly do exist with the representatives
$\int dxF(\Phi,\partial_\mu\Phi)$.

In the present work we find the general solution to the equations

\begin{equation}\label{1.9}
\Delta X=0\quad\hbox{and}\quad\bar{\Delta}^aX=0,
\end{equation}
and also
\begin{equation}\label{1.10}
\Delta e^{{i\over\hbar}S}=0\quad\hbox{and}\quad
\bar{\Delta}^ae^{{i\over\hbar}S}=0,
\end{equation}
for some operators $\Delta$ and $\bar{\Delta}^a$, in case of
finite number of variables. From the field theory point of view, it appears
in all the cases considered that the solutions to the equations (\ref{1.9})
and (\ref{1.10}) are physically trivial. Thus the results obtained, which are
apparently interesting even by themselves, show actually the importance of
the locality condition in quantum field theory. It becomes quite clear that
the locality condition is, in fact, the only requirement which allows one to
obtain physically nontrivial results within the formalism based on the
quantum master equation.

All the equations under consideration are solved in terms of formal power series
expansions with respect to all variables the functions $X$ and $S$ depend on,
and also with respect to the Planck constant entering the $S$ as well.

\section{Equation $\Delta X$ $=$ $0$.}
\setcounter{equation}{0}

Let us denote $\Phi^A=(u^i,\xi^\alpha)$, where $u^i, i=1,\ldots,n_+$ and
$\xi^\alpha,\alpha=1,\ldots,n_-$ are Bosonic and Fermionic variables,
respectively. In this notation the operator $\Delta$ takes the form

\begin{equation}\label{2.1}
\Delta=(-1)^{\varepsilon_A}{\partial\over\partial\Phi^A}
{\partial\over\partial\Phi^*_A}={\partial\over\partial u^i}
{\partial\over\partial u^*_i}-{\partial\over\partial\xi^\alpha}
{\partial\over\partial\xi^*_\alpha},\quad\Delta^2=0,
\end{equation}
where the variables $u^*_i$ ($\xi^*_\alpha$) are Fermionic (Bosonic).

To construct the general solution to the equation

\begin{equation}\label{2.2}
\Delta X=0,
\end{equation}
let us introduce the operator $\Gamma$,

\begin{equation}\label{2.3}
\Gamma=(-1)^{\varepsilon_A}\Phi^A\Phi^*_A=u^iu^*_i-\xi^\alpha\xi^*_\alpha.
\end{equation}
with the properties

\begin{equation}\label{2.3a}
\Gamma\Delta+\Delta\Gamma=N,\quad\Gamma^2=0,\quad[\Delta,N]=[\Gamma,N]=0,
\end{equation}

\begin{equation}\label{2.4}
N=u^i{\partial\over\partial u^i}+
\xi^*_\alpha{\partial\over\partial\xi^*_\alpha}+\left(n-
u^*_i{\partial\over\partial u^*_i}-
\xi^\alpha{\partial\over\partial\xi^\alpha}\right), \quad n=n_++n_-.
\end{equation}
Let us apply the operator $N$ to $X$, and then make use of the equation
(\ref{2.2}):

\begin{equation}\label{2.5}
NX=\Delta\Gamma X.
\end{equation}
If the operator $N$ can be inverted, then the function $X$ can be
represented in the form

\begin{equation}\label{2.6}
X=\Delta Y,\quad Y=\Gamma N^{-1}X.
\end{equation}
Since every Fermionic variable can enter $X$ at most linearly, it follows from
the representation (\ref{2.4})  for $N$ that $NX$ can vanish iff $X$
does not depend on Bosonic variables, while the dependence on Fermionic
variables comes only from their complete product. Thus the general solution
to the equation (\ref{2.2})  can be represented in the form:

\begin{equation}\label{2.7}
X=\Delta Y+c\prod_iu^*_i\prod_\alpha\xi^\alpha,
\end{equation}
where $Y$ is an arbitrary function, and $c$ is an arbitrary constant. It is
of the most importance that the second term in r.h.s. in (\ref{2.7}) cannot be
represented in the form $\Delta Z$. Indeed, otherwise $Z$ would contain $n+1$
Fermionic co--multipliers, which situation would be a contradiction. In other
words one can say that the expression (\ref{2.7}) describes the cohomology
group of the operator $\Delta$ as acting on the space of formal power series
expansions with respect to all variables. The representation (\ref{2.7}) was
independently obtained by K.Bering \cite{12}.

Let us assume, in analogy with field theory,
that original set of variables $\Phi$ is split into two subsets:
$\Phi=(\varphi,c)$, or $u=(\varphi_u,c_\xi)$,
$\xi=(\varphi_\xi,c_u)$,
and the following ghost number values are assigned to all the variables

\begin{equation}\label{2.8}
\hbox{gh}(\varphi_u)=\hbox{gh}(\varphi_\xi)=0, \quad \hbox{gh}(c_\xi)=
\hbox{gh}(c_u)=1,\quad \hbox{gh}(\Phi^*)=-\hbox{gh}(\Phi)-1.
\end{equation}
Then we have

\begin{equation}\label{2.9}
\hbox{gh}\left(\prod_iu^*_i\prod_\alpha\xi^\alpha\right)=n_{c_u}-
n_{\varphi^*_u}-2n_{c^*_\xi}.
\end{equation}
In ``physical'' situation $n_{c_u}<n_{\varphi_u}=n_{\varphi^*_u}$, so that
$\hbox{gh}(\prod u^*_i\prod\xi^\alpha)<0$,
and thus, in the zero ghost number sector, the general solution to the
equation (\ref{2.2}) has the form

\begin{equation}\label{2.10}
X=\Delta Y,\quad \hbox{gh}(X)=0,
\end{equation}
which is physically trivial.

An analogous consideration shows that the general solution to the equation

\begin{equation}\label{2.11}
\Delta_1X=0,\quad\Delta_1\equiv\Gamma,
\end{equation}
has the form

\begin{equation}\label{2.12}
X=\Delta_1Y+c\prod_iu^*_i\prod_\alpha\xi^\alpha.
\end{equation}

\section{Equation $\bar{\Delta}^{a}X = 0$.}
\setcounter{equation}{0}

Let $6n$ be a total number of variables which are, in their own turn, split
into three subsets as follows

\begin{eqnarray}
&&\Phi^A=(u^i,\xi^\alpha),\quad i=1,\ldots,n_+,\quad\alpha=1,\ldots,n_-,\quad
n_++n_-=n, \nonumber\\
&&\bar{\Phi}_A=(\bar{u}_i,\bar{\xi}_\alpha), \nonumber\\
&&\Phi^*_{Aa}=(u^*_{ia},\xi^*_{\alpha a}),\quad
a=1,2,\label{3.1}\\
&&\varepsilon(\Phi^A)=\varepsilon(\bar{\Phi}_A)=\varepsilon_A,\quad
\varepsilon(\Phi^*_{Aa})=\varepsilon_A+1. \nonumber
\end{eqnarray}
Operators $\bar{\Delta}^a$ have the form

\begin{equation}\label{3.2}
\bar{\Delta}^a=(-1)^{\varepsilon_A}{\partial\over\partial\Phi^A}
{\partial\over\partial\Phi^*_{Aa}}+\varepsilon^{ac}\Phi^*_{Ac}
{\partial\over\partial\bar{\Phi}_A}={\partial\over\partial u^i}
{\partial\over\partial u^*_{ia}}-{\partial\over\partial\xi^\alpha}
{\partial\over\partial\xi^*_{\alpha a}}+\varepsilon^{ac}u^*_{ic}
{\partial\over\partial\bar{u}_i}+\varepsilon^{ac}\xi^*_{\alpha c}
{\partial\over\partial\bar{\xi}_\alpha},
\end{equation}
$$
\quad\varepsilon^{ba}=-\varepsilon^{ab},\quad\varepsilon^{12}=-1.
$$
and, thus, are nilpotent

\begin{equation}\label{3.3}
\bar{\Delta}^a\bar{\Delta}^b+\bar{\Delta}^b\bar{\Delta}^a=0,\quad
\varepsilon(\bar{\Delta}^a)=1.
\end{equation}
These operators determine the quantum master equation in $Sp(2)$ symmetric
formulation of gauge theories.

To solve the equation

\begin{equation}\label{3.4}
\bar{\Delta}^aX=0
\end{equation}
let us introduce the operators $\Gamma_a(\theta)$,

\begin{equation}\label{3.5}
\Gamma_a(\theta)=\theta^A\Gamma_{Aa},\quad\Gamma_{Aa}=(-1)^{\varepsilon_A}
(\Phi^*_{Aa}\Phi^A+\varepsilon_{ac}\bar{\Phi}_A
{\partial\over\partial\Phi^*_{Ac}}),
\end{equation}
$$
\varepsilon_{ab}\varepsilon^{bc}=\delta^c_a,\quad
\varepsilon(\theta^A)=0
$$
(there is no summation over $A$ when defining $\Gamma_{Aa}$).
The operators $\Gamma_a(\theta)$ are nilpotent

\begin{equation}\label{3.6}
\Gamma_a(\theta_1)\Gamma_b(\theta_2)+\Gamma_b(\theta_2)\Gamma_a(\theta_1)=0.
\end{equation}
Also, the following relations hold

\begin{eqnarray}
&&\bar{\Delta}^a\Gamma_b(\theta)+\Gamma_b(\theta)\bar{\Delta}^a=
\delta^a_bN(\theta),\nonumber\\
&&[N(\theta),\bar{\Delta}^a]=[N(\theta),\Gamma_a(\theta_1)]=
[N(\theta),N(\theta_1)]=0,\label{3.7}\\
&&N(\theta)=\theta^AN_A,\quad N_i=1+\hat{n}_{u^i}-\hat{n}_{u^*_{i1}}-
\hat{n}_{u^*_{i2}}-\hat{n}_{\bar{u}_i},\quad N_\alpha=1+
\hat{n}_{\xi^*_{\alpha1}}+\hat{n}_{\xi^*_{\alpha2}}+\hat{n}_{\bar{\xi}_\alpha}
-\hat{n}_{\xi^\alpha},
\nonumber
\end{eqnarray}
where $\hat{n}_{u^i}=u^i\partial/\partial u^i$ and so on.
Let us apply the operator $N^2(\theta)$ to the function $X$, and
then make use of the equation (\ref{3.4})

\begin{equation}\label{3.8}
N^2(\theta)X=\bar{\Delta}^2\bar{\Delta}^1\Gamma_1(\theta)\Gamma_2(\theta)X.
\end{equation}
If the operator $N^2(\theta)$ can be inverted for some choice of
parameters $\theta^A$, then $X$ can be represented in the form

\begin{equation}\label{3.9}
X=\bar{\Delta}^2\bar{\Delta}^1Y={1\over2}\varepsilon_{ab}\bar{\Delta}^b
\bar{\Delta}^aY,\quad Y=\Gamma_1(\theta)\Gamma_2(\theta)N^{-2}(\theta)X.
\end{equation}
The operator $N^2(\theta)$ is noninvertible if $X$ satisfies the equations

\begin{equation}\label{3.10}
N_AX=0,\quad\forall A.
\end{equation}
Thus the general solution to the equations (\ref{3.4}) can be represented in
the form

\begin{equation}\label{3.11}
X=\bar{\Delta}^2\bar{\Delta}^1Y+X_BQ,\quad
Q\equiv\prod_\alpha\xi^\alpha,
\end{equation}
where $X_B$ depends only on the variables $u^i$, $u^*_{ia}$, $\bar{u}_i$, and
satisfies the equations

\begin{eqnarray}
&&\bar{\Delta}^a_BX_B=0,\quad(1+\hat{n}_{u^i})X_B=
(\hat{n}_{u^*_{i1}}+\hat{n}_{u^*_{i2}}+\hat{n}_{\bar{u}_i})X_B,
\label{3.12}\\
&&\bar{\Delta}^a_B=\Delta^a_B+d^a,\quad\Delta^a_B={\partial\over\partial u^i}
{\partial\over\partial u^*_{ia}},\quad d^a=\varepsilon^{ac}u^*_{ic}
{\partial\over\partial\bar{u}_i},\nonumber
\end{eqnarray}
(recall that $u^i,\bar{u}_i$ and $u^*_{ia}$ are Bosonic and Fermionic
variables, respectively). Let us expand $X_B$ in power series in $u^*_{ia}$:

\begin{equation}\label{3.13}
X_B=\sum_{k=0}^{2n_+}X^{(k)},\quad(\sum_{i,a}\hat{n}_{u^*_{ia}})X^{(k)}=
kX^{(k)}.
\end{equation}
It follows that $X^{(0)}$ satisfies the equation

\begin{equation}\label{3.14}
\Delta^a_BX^{(0)}=0.
\end{equation}
It is shown in Appendix that the general solution to the equation

\begin{equation}\label{3.15}
\Delta^a_BZ=0
\end{equation}
can be represented in the form

\begin{eqnarray}
&&Z=\Delta^2_B\Delta^1_BZ^\prime+\sum_{n_1+n_2\le n_+}C_{i_1\ldots
i_{n_1}j_1\ldots j_{n_2}}{\partial\over\partial u^*_{i_11}}\ldots
{\partial\over\partial u^*_{i_{n_1}1}}Q_1{\partial\over\partial
u^*_{j_12}}\ldots{\partial\over\partial u^*_{j_{n_2}2}}Q_2,\label{3.16}\\
&&{\partial\over\partial u^*_{ia}}C_{i_1\ldots j_{n_2}}=
{\partial\over\partial u^i}C_{i_1\ldots j_{n_2}}=0,
\quad Q_1\equiv\prod_i u^*_{i1}, \quad Q_2\equiv\prod_i u^*_{i2},\nonumber
\end{eqnarray}
where coefficients $C_{i_1\ldots j_{n_2}}$ are antisymmetric under permutation
of any two neighboring indices.
Thus we have

\begin{equation}\label{3.17}
X^{(0)}=\Delta^2_B\Delta^1_BY^{(2)},
\end{equation}
and $X_B$  rewrites in the form

\begin{eqnarray}
&&X_B=\bar{\Delta}^2_B\bar{\Delta}^1_BY^{(2)}+\sum_{k=1}^{2n_+}X^{(k)}_1,
\nonumber\\
&&\Delta^a_BX^{(1)}_1=0,\quad
X^{(1)}_1=\Delta^2_B\Delta^1_BY^{(3)},\nonumber\\
&&X_B=\bar{\Delta}^2_B\bar{\Delta}^1_B(Y^{(2)}+Y^{(3)})+
\sum_{k=2}^{2n_+}X^{(k)}_2.\label{3.18}
\end{eqnarray}
Continuing this process, we obtain

\begin{eqnarray}
&&X_B=\bar{\Delta}^2_B\bar{\Delta}^1_B(\sum_{k=2}^{n_++1}Y^{(k)})+
\sum_{k=n_+}^{2n_+}X^{(k)}_{n_+},\nonumber\\
&&\Delta^a_BX^{(n_+)}_{n_+}=0\label{3.19}.
\end{eqnarray}
The general solution to the equation (\ref{3.19})) is the following

\begin{equation}\label{3.20}
X^{(n_+)}_{n_+}=\Delta^2_B\Delta^1_BY^{(n_++2)}+\sum_{m=0}^{n_+}a_m(\bar{u})
\varepsilon_{i_1\ldots i_mi_{m+1}\ldots i_{n_+}}
{\partial\over\partial u^*_{i_11}}\ldots {\partial\over\partial u^*_{i_m1}}Q_1
{\partial\over\partial u^*_{i_{m+1}2}}\ldots
{\partial\over\partial u^*_{i_{n_+}2}}Q_2,
\end{equation}
here $\varepsilon_{i_1\ldots i_{n_+}}$ is a totally antisymmetric constant
tensor, and $\varepsilon_{12\ldots n_+}=1$.
The second equation in (\ref{3.12}) yields

\begin{equation}\label{3.21}
{\partial\over\partial \bar{u}_i}a_m(\bar{u})=0,\quad a_m=\hbox{const}.
\end{equation}
Thus we get

\begin{eqnarray}
&&X_B=\bar{\Delta}^2_B\bar{\Delta}^1_B(\sum_{k=2}^{n_++2}Y^{(k)})+\nonumber\\
&&\phantom{11111}\sum_{m=0}^{n_+}a_m
\varepsilon_{i_1\ldots i_mi_{m+1}\ldots i_{n_+}}
{\partial\over\partial u^*_{i_11}}\ldots {\partial\over\partial u^*_{i_m1}}Q_1
{\partial\over\partial u^*_{i_{m+1}2}}\ldots
{\partial\over\partial u^*_{i_{n_+}2}}Q_2
+\sum_{k=n_++1}^{2n_+}X^{(k)}_{n_++1}, \nonumber\\
&&\Delta^a_BX^{(n_++1)}_{n_++1}=0,\label{3.22}\\
&&X^{(n_++1)}_{n_++1}=\Delta^2_B\Delta^1_BY^{(n_++2)}+\sum_{m=0}^{n_+-1}
C_{i_1\ldots i_mi_{m+1}\ldots i_{n_+-1}}
{\partial\over\partial u^*_{i_11}}\ldots {\partial\over\partial u^*_{i_m1}}Q_1
{\partial\over\partial u^*_{i_{m+1}2}}\ldots
{\partial\over\partial u^*_{i_{n_+-1}2}}Q_2 \nonumber.
\end{eqnarray}
Now, the second of the equations (\ref{3.12}) yields

\begin{equation}\label{3.23}
C_{i_1\ldots i_{n_+-1}}=0.
\end{equation}
Continuing this process, we obtain finally

\begin{eqnarray}
&&X_B=\bar{\Delta}^2_B\bar{\Delta}^1_B\tilde{Y}+
\sum_{m=0}^{n_+}a_mF_m,\label{3.24}\\
&&F_m=\varepsilon_{i_1\ldots i_mi_{m+1}\ldots i_{n_+}}
{\partial\over\partial u^*_{i_11}}\ldots {\partial\over\partial u^*_{i_m1}}Q_1
{\partial\over\partial u^*_{i_{m+1}2}}\ldots
{\partial\over\partial u^*_{i_{n_+}2}}Q_2.\nonumber
\end{eqnarray}
Then we find for $X$

\begin{equation}\label{3.25}
X=\bar{\Delta}^2\bar{\Delta}^1Y+\sum_{m=0}^{n_+}a_mF_mQ.
\end{equation}

Let us show that the second term in (\ref{3.25}) represents nontrivial
"cohomologies" of the operators $\bar{\Delta}^a$ in the sense that
there is no combination of functions $F_mQ$, which can be represented in the
form $\bar{\Delta}^2\bar{\Delta}^1Z$.\footnote{Of course, one can represent
the functions $F_mQ$ in the form

$$
F_mQ=\bar{\Delta}^1Z^{(1)}_m=\bar{\Delta}^2Z^{(2)},\quad0<m<n,
$$
$$
F_0Q=\bar{\Delta}^1Z^{(1)}_0,\quad F_nQ=\bar{\Delta}^2Z^{(2)}_n.
$$
}
It is easy to see that the relation

\begin{equation}\label{3.26}
\sum_{m=0}^{n_+}a_mF_mQ=\bar{\Delta}^2\bar{\Delta}^1Z
\end{equation}
implies

\begin{equation}\label{3.27}
\sum_{m=0}^{n_+}a_mF_m=\bar{\Delta}^2_B\bar{\Delta}^1_BZ_B.
\end{equation}
First of all, let us show the relation

\begin{equation}\label{3.28}
cQ_1=\bar{\Delta}^2_B\bar{\Delta}^1_BZ_B,\quad c=\hbox{const},
\end{equation}
to imply $c=0$.
Let us write down explicitly the operator
$\bar{\Delta}^2_B\bar{\Delta}^1_B$ taken at $u^*_{i2}=0$:

\begin{equation}\label{3.29}
\left.\bar{\Delta}^2_B\bar{\Delta}^1_BZ_B\right|_{u^*_{i2}=0}=
\left.{\partial\over\partial u^*_{i1}}\left(u^*_{i1}
{\partial\over\partial \bar{u}_j}{\partial\over\partial u^j}-
{\partial\over\partial u^*_{j2}}{\partial\over\partial u^i}
{\partial\over\partial u^j}\right)Z_B\right|_{u^*_{i2}=0}.
\end{equation}
Thus we see that r.h.s. of the relation (\ref{3.28}) for $u^*_{i2}=0$ can not
contain complete product of the variables $u^*_{i1}$, which fact, in its own
turn, implies $c=0$.

Next, let us introduce the operator

\begin{equation}\label{3.30}
L=u^*_{i1}{\partial\over\partial u^*_{i2}}.
\end{equation}
We shall make use of the following its properties

\begin{eqnarray}
&&[L,\bar{\Delta}^2_B\bar{\Delta}^1_B]=0,\nonumber\\
&&L^mF_m=(-1)^{m(n_+-1)}m!Q_1,\label{3.31}\\
&&L^mF_k=0,\quad k<m.\nonumber
\end{eqnarray}
By applying the operator $L^{n_+}$ to the relation (\ref{3.27}), we get

\begin{equation}\label{3.32}
n_+!a_{n_+}Q_1=\bar{\Delta}^2_B
\bar{\Delta}^1_B(L^{n_+}Z_B),
\end{equation}
so that $a_{n_+}=0$. In the same way, applying the operator $L^{n_+-1}$ to the
relation (\ref{3.27}), we get $a_{n_+-1}=0$, and so on. Finally, we conclude
that the relation (\ref{3.27}) can be fulfilled iff

\begin{equation}\label{3.33}
a_m=0,\quad\forall m.
\end{equation}

So, the general solution to the equation (\ref{3.4}) is given by the
expression (\ref{3.25}), where the second term in r.h.s. of (\ref{3.25})
represents "nontrivial common cohomologies" of the operators
$\bar{\Delta}^a$.

In physical applications, the variables $\Phi$ are split into the subsets
$\Phi$ $=$ $(\varphi,c^a,B)$, so that

\begin{eqnarray}
&&u=(\varphi_u,c^a_\xi,B_u),\quad\xi=(\varphi_\xi,c^a_u,B_\xi), \nonumber\\
&&\bar{u}=(\bar{\varphi}_u,\bar{c}^a_\xi,\bar{B}_u),\quad
\bar{\xi}=(\bar{\varphi}_\xi,\bar{c}^a_u,\bar{B}_\xi), \nonumber \\
&&u^*_a=(\varphi^*_{ua},c^*_{\xi b|a},B^*_{ua}),\quad
\xi^*_a=(\varphi^*_{\xi a},c^*_{ub|a},B^*_{\xi a}), \label{3.34}\\
&&Q=\prod\varphi_\xi\prod c^a_u\prod B_\xi, \nonumber\\
&&Q_1=\prod\varphi^*_{u1}\prod c^*_{\xi b|1}\prod B^*_{u1},\quad
Q_2=\prod\varphi^*_{u2}\prod c^*_{\xi b|2}\prod B^*_{u2}.\nonumber
\end{eqnarray}
The following new ghost number (ngh) values are assigned to all the variables

\begin{eqnarray}
&&\hbox{ngh}(\varphi)=0,\quad\hbox{ngh}(c^a)=1,\quad\hbox{ngh}(B)=2,
\nonumber\\
&&\hbox{ngh}(\Phi^*)=-\hbox{ngh}(\Phi)-1,\quad\hbox{ngh}(\bar{\Phi})=
-\hbox{ngh}(\Phi)-2.\label{3.35}
\end{eqnarray}
Then

\begin{equation}\label{3.36}
\hbox{ngh}(F_mQ)=\hbox{ngh}(Q_1Q)=-n_{\varphi_u}-4n_{c^1_\xi}-3n_{B_u}+
2n_{c^1_u}+2n_{B_\xi}=-n_{\varphi_u}-n_{B_u}-2n_{B_\xi}<0,
\end{equation}
where we have taken into account that $n_{c^1_u}=n_{B_u}$,
$n_{c^1_\xi}=n_{B_\xi}$. Thus, in the zero new ghost
number sector, the equation (\ref{3.4}) has only trivial solutions

\begin{equation}\label{3.37}
X=\bar{\Delta}^2\bar{\Delta}^1Y,\quad\hbox{ngh}(X)=0.
\end{equation}

\section{Equation $\Delta e^{{i\over\hbar}S}=0$}.
\setcounter{equation}{0}

In this Section  we consider the quantum master equation

\begin{equation}\label{4.1}
\Delta e^{{i\over\hbar}S}=0,\quad\varepsilon(S)=0,
\end{equation}
with the operator $\Delta$ defined by the formula (\ref{2.1}). Of course, if
one solves this equation for the exponential in terms of formal power series
expansions, then the general solution is given by the formula (\ref{2.4}),
$X=\exp(iS/\hbar)$. If, however, one requires the function $S$ itself to
expand in formal power series not only with respect to the variables, but
also with respect to the Planck constant $\hbar$, under the extra condition
that the classical limit of $S$ at $\hbar$ $\rightarrow$ $0$ does not vanish,
then the general solution appears to be of quite different structure.  Let us
rewrite the equation (\ref{4.1}) in the following equivalent form

\begin{equation}\label{4.2}
{1\over2}(S,S)=i\hbar\Delta S,
\end{equation}
where $(F,G)$ denotes the antibracket

\begin{equation}\label{4.3}
(F,G)\equiv F{\overleftarrow{\partial}\over\partial\Phi^A}
{\overrightarrow{\partial}\over\partial\Phi^*_A}G-
F{\overleftarrow{\partial}\over\partial\Phi^*_A}
{\overrightarrow{\partial}\over\partial\Phi^A}G.
\end{equation}
Let us expand $S$ in formal power series in $\hbar$:

\begin{equation}\label{4.4}
S=\sum_{k=0}^\infty\hbar^kS^{(k)}.
\end{equation}
Then it follows that $S^{(0)}$ satisfies the classical master equation

\begin{equation}\label{4.5}
(S^{(0)},S^{(0)})=0.
\end{equation}
Let us solve this equation in terms of formal power series expansions with
respect to the variables.
Without any lost of generality one can assume that the expansion of $S^{(0)}$
begins with the contribution quadratic in the variables:

\begin{equation}\label{4.6}
S^{(0)}=\sum_{k=2}^\infty S^{(0)}_k,\quad S^{(0)}_k\sim\sum_{l+m=k}
O(\Phi^l(\Phi^*)^m).
\end{equation}
In its own turn, the lowest order, $S^{(0)}_{2}$, also satisfies the master
equation

\begin{equation}\label{4.7}
(S^{(0)}_2,S^{(0)}_2)=0.
\end{equation}
Let us denote $\eta=(\omega^A;\omega^*_A)$, $\omega^A=(u^i;-\xi^*_\alpha)$,
$\omega^*_A=(u^*_i;\xi^\alpha)$, $\varepsilon(\omega^A)=0$,
$\varepsilon(\omega^*_A)=1$, $A=1,\ldots,n$. Then the equation (\ref{4.7})
takes the form

\begin{equation}\label{4.8}
{\partial S^{(0)}_2\over\partial\omega^A}
{\partial S^{(0)}_2\over\partial\omega^*_A}=0.
\end{equation}
The function $S^{(0)}_{2}$ has the following general structure

\begin{equation}\label{4.9}
S^{(0)}_2={1\over2}\omega^AL_{AB}\omega^B+{1\over2}\omega^*_AM^{AB}\omega^*_B,
\end{equation}
where $L_{AB}$ is a symmetric matrix, while $M^{AB}$ is an antisymmetric one.
Let $O_A^B$ be an orthogonal matrix diagonalizig the matrix $L^{AB}$:

\begin{equation}\label{4.10}
O^C_AO^D_BL_{CD}=\delta_{AB}l_B=\delta_{AB}\zeta_B\lambda^2_B,
\end{equation}
where $\zeta_{B}=\pm1,0$. Let us perform an anticanonical (AC) transformation
(i.e. the one preserving the antibracket (\ref{4.3})) such that

\begin{equation}\label{4.11}
\omega^A=O^A_B\Lambda^{-1}{}^B_C\omega^{\prime C},\quad
\omega^*_A=O^B_A\Lambda^C_B\omega^{*\prime}_C,\quad
\Lambda^A_B=\delta^A_B\lambda_B.
\end{equation}
In new variables (we omit primes), the action $S^{(0)}_2$ takes the form

\begin{equation}\label{4.12}
S^{(0)}_2={1\over2}x^i\zeta_ix^i+{1\over2}x^*_iB^{ij}x^*_j+x^*_iC^{i\alpha}
y^*_\alpha+{1\over2}y^*_\alpha D^{\alpha\beta}y^*_\beta,
\end{equation}
where $\omega^A=(x^i,y^\alpha)$; variables $x^i$ and $y^\alpha$
correspond to $\zeta_i=\pm1$ and $\zeta_\alpha=0$, respectively. By
re-enumerating the variables (which is also an AC transformation),
one can place the values of $\zeta_i$ to take the standard order

\begin{equation}\label{4.13}
\zeta_i=1,\quad i=1,\ldots,m_+,\quad \zeta_i=-1,\quad
i=m_++1,\ldots,m_++m_-.
\end{equation}
Let us substitute the expression (\ref{4.12}) for $S ^{(0)}_{2}$ into the
master equation (\ref{4.8}):

\begin{equation}\label{4.14}
x^i\zeta_iB^{ij}x^*_j+x^i\zeta_iC^{i\alpha}y^*_\alpha=0,
\end{equation}
which yields

\begin{equation}\label{4.15}
B^{ij}=0,\quad C^{i\alpha}=0.
\end{equation}
Further, let us restrict ourselves by considering the (physically interesting)
case of proper solutions (this means that the Hessian of the function
$S^{(0)}_2$ has the rank $n$). For such solutions, the (antisymmetric) matrix
$D^{\alpha \beta}$ is invertible, and, thus, it can be transformed to take
the standard Jordan form, with the help of an orthogonal transformation.
Then, by making use of an AC transformation nontrivial only in the
$y^\alpha, y^*_\alpha$ sector, the action $S ^{(0)}_{2}$ reduces to take
the following canonical form

\begin{equation}\label{4.16}
S^{(0)}_{2,c}={1\over2}x^i\zeta_ix^i+{1\over2}y^*_\alpha
\varepsilon^{\alpha\beta}y^*_\beta,
\end{equation}
where $\varepsilon^{\alpha\beta}$ is a matrix whose Jordan form consists of
the $2\times2$--Jordan blocks equal to $i\sigma^2$.

Let us call two functions $F(\eta)$ and $G(\eta)$ AC equivalent, if
they are connected with each other by an AC transformation: $G(\eta)$ $=$
$F(\eta^\prime(\eta))$, $\eta$ $\rightarrow$ $\eta^\prime$ is AC
transformation. Then
one can say that the general solution to the master equation (\ref{4.8}) is
AC equivalent to the canonical solution of the form (\ref{4.16})
characterized by the pair of negative numbers $m_+$, $m_-$, $m_++m_-\le n$.

Let us turn to solving the
master equation (\ref{4.5}) to higher orders.  We suppose that one has
already applied the anticanonical transformation reducing the $S^{(0)}_{2}$
to the canonical form (\ref{4.16}). Then we get the following equation for
$S^{(0)}_{3}$

\begin{eqnarray}
&&\gamma S^{(0)}_3=0, \label{4.17}  \\
&&\gamma=x^i\zeta_i{\partial\over\partial x^*_i}-
y^*_\alpha\varepsilon^{\alpha\beta}{\partial\over\partial y^\beta},\quad
\gamma^2=0,\nonumber
\end{eqnarray}
Introduce the operator $d$,

\begin{equation}\label{4.18}
d=x^*_i\zeta_i{\partial\over\partial x^i}-
y^\alpha\varepsilon_{\alpha\beta}{\partial\over\partial y^*_\beta},\quad
\varepsilon_{\alpha\beta}\varepsilon^{\beta\gamma}=\delta^\gamma_\alpha,
\end{equation}
with the properties

\begin{eqnarray}
&&d^2=0,\quad d\gamma+\gamma d=x^i{\partial\over\partial x^i}+
y^\alpha{\partial\over\partial y^\alpha}+
x^*_i{\partial\over\partial x^*_i}+
y^*_\alpha{\partial\over\partial y^*_\alpha}\equiv N,\label{4.19}\\
&&[N,d]=[N,\gamma]=0.\nonumber
\end{eqnarray}
By making use of the standard reasoning analogous to the one given in Section
2, we find that the general solution to the equation

\begin{equation}\label{4.20}
\gamma X=0
\end{equation}
has the form

\begin{equation}\label{4.21}
X=\gamma Y+ c,\quad c=\hbox{const}.
\end{equation}

Now, returning to the equation (\ref{4.17}), we conclude that the
$S^{(0)}_{3}$ can be represented in the form

\begin{equation}\label{4.22}
S^{(0)}_3=\gamma Y_3=-(Y_3,S^{(0)}_{2,c}),\quad\varepsilon(Y_3)=1.
\end{equation}
Let us introduce the function $S^{\prime (0)}$ AC equivalent to the one
$S^{(0)}$:

\begin{equation}\label{4.23}
S^{\prime(0)}=e^{\hat{Y}_3}S^{(0)},\quad\hat{Y}F\equiv(Y,F).
\end{equation}
Then the $S^{\prime (0)}$ expands in power series with respect to the
variables $\eta$, in the form

\begin{equation}\label{4.24}
S^{\prime(0)}=S^{(0)}_{2,c}+S^{\prime(0)}_4 +O(\eta^5),
\end{equation}
and $S^{\prime (0)}_{4}$ satisfies the equation

\begin{equation}\label{4.25}
\gamma S^{\prime(0)}_4=0,\quad S^{\prime(0)}_4=\gamma
Y_4=-(Y_4,S^{(0)}_{2,c}).
\end{equation}
Continuing this process, we conclude that the general proper solution to the
classical master equation (\ref{4.5}) is AC equivalent to the
$S^{(0)}_{2,c}$:

\begin{equation}\label{4.26}
S^{(0)}=e^{\hat{Y}^{(0)}}S^{(0)}_{2,c},\quad\varepsilon(Y^{(0)})=1.
\end{equation}

Before we turn to the quantum master equation, let us introduce the operators
we call "gauge transformations" (G transformations)

\begin{eqnarray}
&&K=e^{[\Delta,Z]_+}=e^\rho e^{-\hat{Z}},\quad\varepsilon(Z)=1,\label{4.27}\\
&&\rho=\ln{D(\Theta)\over D(\eta)},\quad\Theta=e^{-\hat{Z}}\eta.\nonumber
\end{eqnarray}
We shall make use of the following properties of G transformations.

Given the
function $S(\eta,\hbar)$, let us construct the following
$S^{\prime}(\eta,\hbar)$:

\begin{equation}\label{4.28}
e^{{i\over\hbar}S^\prime(\eta,\hbar)}=
e^{[\Delta,Z]_+}e^{{i\over\hbar}S(\eta,\hbar)}.
\end{equation}
Then we have

\begin{eqnarray}
&&S^{\prime(0)}(\eta)=S^\prime(\eta,0)=e^{-\hat{Z}^{(0)}}S(\eta,0)=
e^{-\hat{Z}^{(0)}}S^{(0)}(\eta)\label{4.29}\\
&&\Delta e^{{i\over\hbar}S(\eta,\hbar)}=0\quad\Rightarrow\quad
\Delta e^{{i\over\hbar}S^\prime(\eta,\hbar)}=0.\nonumber
\end{eqnarray}
Besides, being applied successively, the G transformations result in complete
G transformation, again.

To construct a solution to the quantum master equation (\ref{4.1}), let us
apply the following G transformation to the action $S$:

\begin{equation}\label{4.30}
e^{{i\over\hbar}S^\prime}=e^{[\Delta,Y^{(0)}]_+}e^{{i\over\hbar}S}.
\end{equation}
Then $S^{\prime(0)}=S^{(0)}_{2,c}$, and, by making use of the property
$\Delta S^{(0)}_{2,c}=0$, we find:

\begin{equation}\label{4.31}
\gamma S^{\prime(1)}=0,\quad S^{\prime(1)}=\gamma Y^{(1)}+c^{(1)}=-
(Y^{(1)},S^{(0)}_{2,c})+c^{(1)}.
\end{equation}
Let us introduce the action $S^{\prime\prime}$,

\begin{equation}\label{4.32}
e^{{i\over\hbar}S^{\prime\prime}}=e^{-\hbar[\Delta,Y^{(1)}]}
e^{{i\over\hbar}S^\prime}.
\end{equation}
whose $\hbar$--power series expansion is

\begin{equation}\label{4.33}
S^{\prime\prime}=S^{(0)}_{2,c}+\hbar c^{(1)}+\hbar^2S^{\prime\prime(2)}+
O(\hbar^3).
\end{equation}
It follows from the quantum master equation (\ref{4.1}) that

\begin{equation}\label{4.34}
\gamma S^{\prime\prime(2)}=0,\quad S^{\prime\prime(2)}=
\gamma Y^{(2)}+c^{(2)}=- (Y^{(2)},S^{(0)}_{2,c})+c^{(2)}.
\end{equation}

Continuing this process, we conclude that the general proper solution to the
quantum master equation (\ref{4.1}) is a result of a G transformation applied
to the $S^{(0)} _{2,c}$:

\begin{equation}\label{4.35}
e^{{i\over\hbar}S}=e^{[\Delta,Y]_+}e^{{i\over\hbar}(S^{(0)}_{2,c}+c)}.
\end{equation}

Would an analogous result be valid in field theory, this would imply physical
triviality of all the field--theoretic models. The locality condition is the
only one which causes the existence of physically nontrivial nonequivalent
theories.

\section{Equation $\bar{\Delta}^a_\hbar e^{{i\over\hbar}S}=0$}.
\setcounter{equation}{0}

To begin with, let us make a few preliminary remarks. We suppose that the
variables are split into the following subsets

\begin{equation}\label{5.1}
\Phi=(\varphi^I,c^{\alpha b},B^\alpha),\quad \bar{\Phi}=(\bar{\varphi}_I,
\bar{c}_{\alpha b},\bar{B}_\alpha),\quad\Phi^*_a=(\varphi^*_{Ia},
c^*_{\alpha b|a},B^*_{\alpha a}).
\end{equation}
The new ghost number values are assigned to all the variables, in
the same way as in Section 3, eqs (\ref{3.35}).
The group $Sp(2)$ is supposed to act on the space of all variables (in fact,
the $Sp(2)$ applies to the indices $a,b$). The group is realized as a set of
linear changes of variables. We assume that the variables $\varphi$, $B$,
$\bar{\varphi}$, $\bar{B}$ are $Sp(2)$ scalars, the $c^a$ form doublet
representations, the $\bar{c}_a$, $\varphi^*_a$, $B^*_a$ form antidoublet representations,
and the $c^*_{b|a}$ transform as a product of two antidoublets.

The operators
$\bar{\Delta}^a_\hbar$ have the form

\begin{equation}\label{5.2}
\bar{\Delta}^a_\hbar=(-1)^{\varepsilon(\Phi)}{\partial\over\partial\Phi}
{\partial\over\partial\Phi^*_a}+{i\over\hbar}\varepsilon^{ac}\Phi^*_c
{\partial\over\partial\bar{\Phi}}, \quad \hbox{ngh}(\bar{\Delta}^a_\hbar)=1.
\end{equation}
They are nilpotent

\begin{equation}\label{5.3}
\bar{\Delta}^a_\hbar\bar{\Delta}^b_\hbar+
\bar{\Delta}^b_\hbar\bar{\Delta}^a_\hbar=0
\end{equation}
and form an $Sp(2)$ doublet.

We will study the general solution to the equation

\begin{equation}\label{5.4}
\bar{\Delta}^a_\hbar e^{{i\over\hbar}S}=0,
\end{equation}
which appears naturally in $Sp(2)$ symmetric formulation of quantized gauge
theories, and which we will call $Sp(2)$ quantum master equation ($Sp(2)$
QME).

Let us introduce the class of operators which we call gauge (G)
transformations

\begin{equation}\label{5.5}
K=e^{{1\over2}\varepsilon_{ab}[\bar{\Delta}^b_\hbar,[\bar{\Delta}^a_\hbar,
F]]_+},
\end{equation}
where $F$ is an arbitrary function or differential operator. If $F$ is an
$Sp(2)$ scalar, then $K$ does the same. If $\hbox{ngh}(F)=-2$, then
$\hbox{ngh}(K)=0$ (in this case we say that the operator $K$ preserves
$\hbox{ngh}$).  If $S$ satisfies $Sp(2)$ QME (\ref{5.4}), then $S^\prime$
defined by the formula

\begin{equation}\label{5.6}
e^{{i\over\hbar}S^\prime}=Ke^{{i\over\hbar}S},
\end{equation}
does the same.
It is easy to check that G transformations applied successively result,
again, in a complete G transformation. Besides, as it was shown in \cite{13},
an arbitrary change of variables, $\Phi$ $\rightarrow$ $\Phi^\prime=F(\Phi)$,
can be extended to become a G transformation, in the following sense.
Let us represent the $F(\Phi)$ in the form

\begin{equation}\label{5.7}
F^A(\Phi)=e^{f^C(\Phi){\partial\over\partial\Phi^C}}\Phi^A,
\end{equation}
with some functions $f^C(\Phi)$ (such a representation is always possible
perturbatively with respect to the variables). Let us consider a G
transformation of the form

\begin{equation}\label{5.8}
K_f=e^{{i\hbar\over2}\varepsilon_{ab}[\bar{\Delta}^b_\hbar,
[\bar{\Delta}^a_\hbar,\bar{\Phi}_Af^A]]_+}.
\end{equation}
Then the relation holds

\begin{equation}\label{5.9}
\left.K_fe^{{i\over\hbar}S(\Phi,\Phi^*_a,\bar{\Phi})}
\right|_{\Phi^*_a=\bar{\Phi}=0}=e^{{i\over\hbar}[S(F(\Phi),0,0)+O(\hbar)]}.
\end{equation}
Moreover, if one deals with the linear change of variables

\begin{equation}\label{5.10}
f^C(\Phi)=f^C_D\Phi^D,
\end{equation}
then the operator $K_{f}$ also reduces, in essential, to the linear
transformation

\begin{eqnarray}
&&S^\prime=e^{U_f}S+\hbar(-1)^{\varepsilon_A}f^A_A, \label{5.11}\\
&&U_f=f^C_D\Phi^D{\partial\over\partial\Phi^C}-
f^D_C\bar{\Phi}_D{\partial\over\partial\bar{\Phi}_C}-
f^D_C\Phi^*_{Da}{\partial\over\partial\Phi^*_{Ca}}.\nonumber
\end{eqnarray}
If the change $\Phi$ $\rightarrow$ $\Phi^\prime$ is $Sp(2)$ covariant and
preserves $\hbox{ngh}$, then the operator $K_{f}$ does the same.

Now, let us turn to solving of the equation (\ref{5.4}). We assume that the
$S$ is a Boson, $\varepsilon(S)=0$, preserves $\hbox{ngh}$:
$\hbox{ngh}(S)=0$, and is an $Sp(2)$ scalar.  Let us seek for a solution in
terms of power series expansion with respect to the variables and Planck
constant $\hbar$ as well:

\begin{equation}\label{5.12}
S=\sum_{k=0}^\infty\hbar^kS^{(k)}.
\end{equation}
The $S^{(0)}$ satisfies the $Sp(2)$ master equation (Sp(2) ME)

\begin{eqnarray}
&&{1\over2}(S^{(0)},S^{(0)})^a+\varepsilon^{ac}\Phi^*_{Ac}
{\partial\over\partial\bar{\Phi}_A}S^{(0)}=0,\label{5.13}\\
&&(F,G)^a\equiv F{\overleftarrow{\partial}\over\partial\Phi^A}
{\overrightarrow{\partial}\over\partial\Phi^*_{Aa}}G-
F{\overleftarrow{\partial}\over\partial\Phi^*_{Aa}}
{\overrightarrow{\partial}\over\partial\Phi^A}G.\nonumber
\end{eqnarray}
Let us expand the $S^{(0)}$ in power series with respect to the variables

\begin{equation}\label{5.14}
S^{(0)}=\sum_{k=2}^\infty S^{(0)}_k,\quad S^{(0)}_k\sim \Gamma^k,
\end{equation}
where $\Gamma$ denotes complete set of variables. The $S^{(0)}_2$ itself
satisfies $Sp(2)$ ME

\begin{equation}\label{5.15}
{1\over2}(S^{(0)}_2,S^{(0)}_2)^a+\varepsilon^{ac}\Phi^*_{Ac}
{\partial\over\partial\bar{\Phi}_A}S^{(0)}_2=0.
\end{equation}
The general structure of $S^{0}_{2}$, which preserves $\hbox{ngh}$ and is an
$Sp(2)$ scalar, is given by the formula

\begin{equation}\label{5.16}
S^{(0)}_2=\varphi^ID_{1II^\prime}\varphi^{I^\prime}+
\varphi^*_{Ia}D^I_{2\alpha}c^{\alpha a}+
\bar{\varphi}_ID^I_{3\alpha}B^\alpha+
\varepsilon^{ab}c^*_{\alpha b|a}D^\alpha_{4\beta}B^\beta.
\end{equation}
With the help of a linear transformation of the variables $\varphi$ , in the
same way as described in Section 4, one can reduce the matrix $D_1$ to take
the standard form

\begin{equation}\label{5.17}
\varphi^ID_{1II^\prime}\varphi^{I^\prime}\quad\rightarrow\quad x^i\Lambda^{ij}
x^j,
\end{equation}
where $\Lambda_{ij}$ is diagonal in the sector of Bosonic variables, with
$m_+$ eigenvalues $+1$ and $m_-$ eigenvalues $-1$, and has the Jordan form in
the sector of Fermionic variables, with $m$ Jordan blocks $i\sigma^2$,
$\varphi^I=(x^i,y^\mu)$, $y^\mu$ are variables complement to $x^i$ (the
kernel of the matrix $D_1$).  As a linear transformation of the variables
$\varphi$ is extendable to become a linear transformation of all variables,
which is, in turn, a G transformation preserving $Sp(2)$ and $\hbox{ngh}$, we
can suppose that we have it already applied, so that the $S^{(0)}_{2}$ takes
the form

\begin{equation}\label{5.18}
S^{(0)}_2={1\over2}x^i\Lambda_{ij}x^j+x^*_{ia}D^i_{1\alpha}c^{\alpha a}+
y^*_{\mu a}D^\mu_{2\alpha}c^{\alpha a}+\bar{x}_iD^i_{3\alpha}B^\alpha+
\bar{y}_\mu D^\mu_{4\alpha}B^\alpha-\varepsilon^{ab}c^*_{\alpha a|b}
D^\alpha_{5\beta}B^\beta.
\end{equation}
By substituting the expression (\ref{5.18}) into the $Sp(2)$ ME (\ref{5.15}),
we get

\begin{equation}\label{5.19}
D^j_{1\alpha}=0,\quad D^\mu_{4\alpha}=D^\mu_{2\beta}D^\beta_{5\alpha}.
\end{equation}
Note that the action $S$ ($S^{(0)}$) for $\bar{\varphi}=\varphi^*_2=0$
satisfies the quantum master equation (master equation) in the variables
$\varphi,\varphi^*_1$, while for $\bar{\varphi}=\varphi^*_1=0$ the actions do
the same in the variables $\varphi,\varphi^*_2$.
In both cases we suppose these solutions to be proper, which means that
$D^\mu_{2\alpha}$ and $D^\mu_{4\alpha}$ are square invertible matrices.  The
$S^{(0)}_{2}$ is of the form

\begin{equation}\label{5.20}
S^{(0)}_2={1\over2}x^i\Lambda_{ij}x^j+y^*_{\alpha a}D^\alpha_{2\beta}
c^{\beta a}+\bar{y}_\alpha D^\alpha_{4\beta}B^\beta+\varepsilon^{ab}
c^*_{\alpha a|b}D^{-1\alpha}_{2\gamma}D^\gamma_{4\beta}B^\beta.
\end{equation}
Let us transform the action $S$ by applying the G transformation
corresponding to the following linear change

\begin{equation}\label{5.21}
c^{\alpha a}\quad\rightarrow\quad c^{\prime\alpha a}=D^\alpha_{2\beta}
c^{\beta a},\quad B^\alpha\quad\rightarrow\quad B^{\prime\alpha}=
D^\alpha_{4\beta}B^\beta,
\end{equation}
which reduces the action $S^{(0)}_2$ to take the canonical form

\begin{equation}\label{5.22}
S^{(0)}_{2,c}={1\over2}x^i\Lambda_{ij}x^j+y^*_{\alpha a}c^{\alpha a}+
\bar{y}_\alpha B^\alpha+\varepsilon^{ab}c^*_{\alpha a|b}B^\alpha.
\end{equation}
The contribution $S^{(0)}_{3}$ satisfies the equation

\begin{eqnarray}
&&(S^{(0)}_2,S^{(0)}_3)^a\equiv\gamma^aS^{(0)}_3=0,\label{5.23}\\
&&\gamma^a=x^i\Lambda_{ij}{\partial\over\partial x^*_{ja}}-
(-1)^{\varepsilon(c^{\alpha a})}c^{\alpha a}{\partial\over\partial y^\alpha}+
y^*_{\alpha b}{\partial\over\partial c^*_{\alpha b|a}}+
(-1)^{\varepsilon(B^\alpha)}\varepsilon^{ab}B^\alpha
{\partial\over\partial c^{\alpha b}}+
\nonumber\\
&&\phantom{11111}(\bar{y}_\alpha+
\varepsilon^{bc}c^*_{\alpha b|c}){\partial\over\partial B^*_{\alpha a}}+
\varepsilon^{ab}(x^*_{ib}{\partial\over\partial\bar{x}_i}+y^*_{\alpha b}
{\partial\over\partial\bar{y}_\alpha}+c^*_{\alpha c|b}
{\partial\over\partial\bar{c}_{\alpha c}}+B^*_{\alpha b}
{\partial\over\partial\bar{B}_\alpha}),\nonumber\\
&&\gamma^a\gamma^b+\gamma^b\gamma^a=0.\nonumber
\end{eqnarray}

To construct a solution to the equation (\ref{5.23}), let us introduce the
operator--valued antidoublet $d_{a}$, $\hbox{ngh}(d_{a}) = -1$,

\begin{equation}\label{5.24}
d_a=x^*_{ia}\Lambda^{-1}{}^{ij}{\partial\over\partial x^j}-
(-1)^{\varepsilon(c^{\alpha a})}y^\alpha{\partial\over\partial c^{\alpha a}}-
(-1)^{\varepsilon(B^\alpha)}\varepsilon_{ab}c^{\alpha b}
{\partial\over\partial B^\alpha}+\varepsilon_{ab}\bar{x}_i
{\partial\over\partial x^*_{ib}}
\end{equation}
with the properties

\begin{eqnarray}
&&d_ad_b+d_bd_a=0,\quad d_a\gamma^b+\gamma^bd_a=\delta^b_aN,\nonumber\\
&&[N,d_a]=[N,\gamma^a]=0,\label{5.25}\\
&&N=x^i{\partial\over\partial x^i}+y^\alpha{\partial\over\partial y^\alpha}+
c^{\alpha a}{\partial\over\partial c^{\alpha a}}+
B^\alpha{\partial\over\partial B^\alpha}+
x^*_{ia}{\partial\over\partial x^*_{ia}}+
\bar{x}_i{\partial\over\partial\bar{x}_i},   \nonumber
\end{eqnarray}
where the operator $N$ is an $Sp(2)$ scalar.

The general solution to the equation

\begin{equation}\label{5.26}
\gamma^aX=0,\quad \hbox{ngh}(X)=0,
\end{equation}
has the form

\begin{equation}\label{5.27}
X={1\over2}\varepsilon_{ab}\gamma^b\gamma^aY+c,\quad c=\hbox{const},\quad
\hbox{ngh}(Y)=-2,
\end{equation}
due to the relation

\begin{equation}\label{5.28}
N^2X={1\over4}\varepsilon_{ab}\gamma^b\gamma^a\varepsilon^{cd}d_dd_cX.
\end{equation}
Note that, being $X$ an $Sp(2)$ scalar, one can choose $Y$ to do the same, as
it follows from (\ref{5.28}).

Thus we have

\begin{equation}\label{5.29}
S^{(0)}_3={1\over2}\varepsilon_{ab}\gamma^b\gamma^aY_3={1\over2}
\varepsilon_{ab}(S^{(0)}_2,(S^{(0)}_2,Y_3)^a)^b,\quad\hbox{ngh}(Y_3)=-2,
\end{equation}
the $Y_3$ is $Sp(2)$ scalar. Let us apply the following G transformation to
$S$

\begin{equation}\label{5.30}
e^{{i\over\hbar}S^\prime}=e^{-{i\hbar\over2}\varepsilon_{ab}[\bar{\Delta}^b,
[\bar{\Delta}^a,Y_3]]_+}e^{{i\over\hbar}S}.
\end{equation}
Then the $S^{\prime (0)}$ expands as

\begin{eqnarray}
&&S^{\prime(0)}=S^{(0)}_{2,c}+S^{(0)}_4+O(\Gamma^5), \nonumber\\
&&\gamma^aS^{(0)}_4=0, \label{5.31}
\end{eqnarray}
and so on. Thus, as a result of having applied a series of successive G
transformations, the general solution $S$ takes the form

\begin{eqnarray}
&&S=S^{(0)}_{2,c}+\hbar S^{(1)}+O(\hbar^2), \nonumber\\
&&\gamma^aS^{(1)}=i\Delta^aS^{(0)}_{2,c}=0,\quad
\Delta^a=(-1)^{\varepsilon_A}{\partial\over\partial\Phi^A}
{\partial\over\partial\Phi^*_{Aa}}. \label{5.32}
\end{eqnarray}

Continuing this reasoning, we conclude that the general solution to the
$Sp(2)$ QME (\ref{5.4}), preserving $Sp(2)$ and $\hbox{ngh}$, is
G equivalent to the canonical solution

\begin{equation}\label{5.33}
e^{{i\over\hbar}S}=e^{{i\hbar\over2}\varepsilon_{ab}[\bar{\Delta}^b,
[\bar{\Delta}^a,Y]]_+}e^{{i\over\hbar}(S^{(0)}_{2,c}+c)},
\end{equation}
where $Y$ is an arbitrary function.

So, one can say again that nontrivial physical result is possible in field
theory only because of the locality condition.

{\bf Acknowledgement}:

The work of I.A.B. and I.V.T. is partially supported by grants
INTAS--RFBR 95--0829i and INTAS 96--0308. The work of I.A.B. is also
supported by grant RFBR 96--01--00482. The work of I.V.T. is also supported
by grant RFBR 96--02--17314.

\def\theequation{A.\arabic{equation}}
\setcounter{equation}{0}

\section*{Appendix}

In this Appendix we consider briefly how to construct the general solution to
the equation

\begin{eqnarray}
&&\Delta^a_BX_B=0,\label{A.1}\\
&&\Delta^a_B={\partial\over\partial u^i}{\partial\over\partial u^*_{ia}},
\quad \varepsilon(u)=0,\quad\varepsilon(u^*)=1,
\quad i=1,\ldots,n.\nonumber
\end{eqnarray}
In what follows we omit the subscript ``$B$''.

As a preliminary step, let us construct the general solution to the equation

\begin{equation}\label{A.2}
\Delta^2\Delta^1U=0.
\end{equation}
Let us expand $U$ in power series with respect to $u^*_{ia}$:

\begin{equation}\label{A.3}
U=\sum_{m,k=0}^nU_{m,k},\quad U_{m,k}\sim(u^*_1)^m(u^*_2)^k.
\end{equation}
It is obvious that each order contribution $U_{m,n}$ satisfies separately the
equation (\ref{A.2}):

\begin{equation}\label{A.4}
\Delta^2\Delta^1U_{m,k}=0.
\end{equation}
Let $m=k=n$:

\begin{equation}\label{A.5}
U_{n,n}=c(u)Q_1Q_2,\quad Q_1\equiv\prod u^*_1,\quad Q_2\equiv\prod u^*_2.
\end{equation}
The equation (\ref{A.4}) yields

\begin{equation}\label{A.6}
c(u)=c+c_iu^i.
\end{equation}
Now, let $m$ or $k$ (or both of them) be less than $n$. Let, for example,
$k<n$.  It follows from the results of Section 2 that

\begin{equation} \label{A.7}
\begin{array}{ll}
\Delta^1U_{m,k}=\Delta^2Y_{m-1,k+1}, &\Delta^2\Delta^1Y_{m-1,k+1}=0 \\
\Delta^1Y_{m-1,k+1}=\Delta^2Y_{m-2,k+2}, &\Delta^2\Delta^1Y_{m-2,k+2}=0 \\
...& \\
\Delta^1Y_{m-l,k+l}=\Delta^2Y_{m-l-1,k+l+1}. &
\end{array}
\end{equation}
Our further actions depend on which of the two equalities, $m-l-1=0$ or
$k+l+1=n$, is first satisfied.

Let $m+k\le n$. Then , for $l=m-1$ in (\ref{A.7}), we get

\begin{equation}\label{A.8}
\Delta^1Y_{1,k+m-1}=\Delta^2Y_{0,k+m}.
\end{equation}
It follows from the results of Section 2 that

\begin{equation}\label{A.9}
Y_{0,k+m}=-\Delta^1Z_{1,k+m},
\end{equation}
which yields

\begin{eqnarray}
&&\Delta^1(Y_{1,k+m-1}-\Delta^2Z_{1,k+m})=0,\nonumber\\
&&Y_{1,k+m-1}=\Delta^2Z_{1,k+m}-\Delta^1Z_{2,k+m-1},\label{A.10}\\
&&\Delta^1(Y_{2,k+m-2}-\Delta^2Z_{2,k+m-1})=0,\nonumber
\end{eqnarray}
and so on. Finally, we obtain

\begin{eqnarray}
&&\Delta^1(U_{m,k}-\Delta^2Z_{m,k+1})=0,\nonumber\\
&&U_{m,k}=\Delta^2Z_{m,k+1}+\Delta^1Z_{m+1,k}\label{A.11}
\end{eqnarray}
(this representation is also valid for $m=n$, $k=0$).

Let $m+k>n$. Then, for $l=n-k-1$ in (\ref{A.7}), we get

\begin{equation}\label{A.12}
\Delta^1Y_{m+k+1-n,n-1}=\Delta^2Y_{m+k-n,n},\quad
\Delta^2\Delta^1Y_{m+k-n,n}=0,
\end{equation}
which yields

\begin{equation}\label{A.13}
\Delta^1Y_{m+k-n,n}=c(u^*_1)Q_2,\quad
c(u^*_1)\sim(u^*_1)^{m+k-1-n}.
\end{equation}
It is convenient to represent the coefficient function $c(u^*_1)$ in the
form

\begin{equation}\label{A.14}
c(u^*_1)=C_{i_1\ldots i_{2n-m-k+1}}{\partial\over\partial u^*_{i_11}}
\ldots{\partial\over\partial u^*_{i_{2n-m-k+1}1}}Q_1,
\end{equation}
with $C_{i_1 ...}$ being totally antisymmetric in its indices.

Let us introduce the following notations

\begin{eqnarray}
&&C_{(m)(k)}\equiv C_{i_1\ldots i_mj_1\ldots j_k},\label{A.15}\\
&&({\partial\over\partial u^*_1})^mQ_1\equiv
{\partial\over\partial u^*_{i_11}}\ldots
{\partial\over\partial u^*_{i_m1}}Q_1,\quad
({\partial\over\partial u^*_2})^kQ_2\equiv
{\partial\over\partial u^*_{j_12}}\ldots
{\partial\over\partial u^*_{j_k2}}Q_2. \nonumber
\end{eqnarray}
Then

\begin{equation}\label{A.16}
c(u^*_1)=C_{(2n-m-k+1)}
({\partial\over\partial u^*_1})^{2n-m-k+1}Q_1.
\end{equation}

The general solution to the equation (\ref{A.13}) for $Y_{m + k -n, n}$ is

\begin{equation}\label{A.17}
Y_{m+k-n,n}=-\Delta^1Z_{m+k+1-n,n}+u^iC_{i(2n-m-k)}
({\partial\over\partial u^*_1})^{2n-m-k+1}Q_1Q_2.
\end{equation}
Then we get for $Y_{m + k + 1 -n, n - 1}$

\begin{equation}\label{A.18}
\Delta^1(Y_{m+k+1-n,n-1}-\Delta^2Z_{m+k+1-n,n})=\xi_1C_{(2n-m-k)(1)}
({\partial\over\partial u^*_1})^{2n-m-k}Q_1
({\partial\over\partial u^*_2})^1Q_2.
\end{equation}
Here $\xi_{1} =\pm1$ is a sign factor which appears when commuting
$\partial/\partial u^*_2$ and $u^*_1$.  Its precise value is
inessential to us.  It follows from (\ref{A.18}) that

\begin{equation}\label{A.19}
Y_{m+k+1-n,n-1}=\Delta^2Z_{m+k+1-n,n}-\Delta^1Z_{m+k+2-n,n-1}+
\xi_1u^iC_{i(2n-m-k-1)(1)}({\partial\over\partial u^*_1})^{2n-m-k-1}Q_1
({\partial\over\partial u^*_2})^1Q_2,
\end{equation}
and so on. Finally, we obtain

\begin{equation}\label{A.20}
U_{m,k}=\Delta^2Z_{m,k+1}+\Delta^1Z_{m+1,k}+
u^iC_{i(n-m)(n-k)}({\partial\over\partial u^*_1})^{n-m}Q_1
({\partial\over\partial u^*_2})^{n-k}Q_2,
\end{equation}
where all sign factors are absorbed into $C_{(2n-m-k+1)}$.

Thus the general solution to the equation (\ref{A.2}) can be represented in
the form

\begin{equation}\label{A.21}
U=\Delta^1Z_1+\Delta^2Z_2+\sum_{m+k<n}
u^iC^{mk}_{i(m)(k)}({\partial\over\partial u^*_1})^mQ_1
({\partial\over\partial u^*_2})^kQ_2.
\end{equation}

Now let us return to solving of the equation (\ref{A.1}). Setting $a=1$, we
find

\begin{equation}\label{A.22}
X=\Delta^1U+c(u^*_2)Q_1.
\end{equation}
The equation (\ref{A.1}) at $a=2$ yields

\begin{equation}\label{A.23}
\Delta^2\Delta^1U=0.
\end{equation}
Substituting the expression (\ref{A.21}) for $U$ into (\ref{A.22}) we obtain

\begin{equation}\label{A.24}
X=\Delta^2\Delta^1Z+\sum_{m+k<n}
C^{mk}_{(m)(k)}({\partial\over\partial u^*_1})^mQ_1
({\partial\over\partial u^*_2})^kQ_2.
\end{equation}

Let us show that the second term in r.h.s. of (\ref{A.24}) cannot be
represented in the form $\Delta^2\Delta^1Z$. Assume that

\begin{equation}\label{A.25}
\sum_{m+k<n}C^{mk}_{(m)(k)}({\partial\over\partial u^*_1})^mQ_1
({\partial\over\partial u^*_2})^kQ_2=\Delta^2\Delta^1Z
\end{equation}
with some coefficients $C^{mk}_{(m+k)}$. Let $m_0$ be the maximal of $m$
entering l.h.s.  of (\ref{A.25}). Introduce the operator $L$,

\begin{equation}\label{A.26}
L=u^*_{i1}{\partial\over\partial u^*_{i2}},
\end{equation}
with the properties

\begin{eqnarray}
&&[L,\Delta^2\Delta^1]=0,\label{A.27}\\
&&LC^{mk}_{(m)(k)}({\partial\over\partial u^*_1})^mQ_1
({\partial\over\partial u^*_2})^kQ_2=(-1)^{n-1}m
C^{mk}_{(m-1)(k+1)}({\partial\over\partial u^*_1})^{m-1}Q_1
({\partial\over\partial u^*_2})^{k+1}Q_2.\nonumber
\end{eqnarray}
By applying the operator $L^{m_0}$ to the relation (\ref{A.25}) we get

\begin{equation}\label{A.28}
\sum_{m_0+k<n}C^{m_0k}_{(m_0+k)}Q_1({\partial\over\partial u^*_2})^{m_0+k}Q_2
={(-1)^{m_0(n-1)}\over m_0!}\Delta^2\Delta^1(L^{m_0}Z),
\end{equation}
which implies $C^{m_0k}_{(m_0+k)}=0$ for all $k$. Thus we
conclude that the relation (\ref{A.25}) can be fulfilled iff
$C^{mk}_{(m+k)}=0$ for all $m$, $k$.

\end{document}